\documentclass[]{spie}  

\usepackage{amsmath,amsfonts,amssymb}
\usepackage{graphicx}
\usepackage[colorlinks=true, allcolors=blue]{hyperref}
\usepackage{glossaries}
\usepackage{listings}       
\usepackage{subcaption}
\usepackage{soul}
\usepackage[normalem]{ulem}
\usepackage{booktabs}
\usepackage{enumitem}

\title{Accelerated Modeling of Near and Far-Field Diffraction for Coronagraphic Optical Systems}

\author[a]{Ewan S. Douglas}
\author[b]{Marshall D. Perrin}
\affil[a]{Massachusetts Institute of Technology, Department of Aeronautics and Astronautics, Cambridge, MA, USA}
\affil[b]{Space Telescope Science Institute, Baltimore, MD, USA}

\authorinfo{Further author information: 
\\E.S.D.: E-mail: douglase@bu.edu,\\  M.D.P: E-mail:  mperrin@stsci.edu}

\newacronym{AU}{AU}{Astronomical Unit [1.5e11 m]}  
\newacronym{pc}{pc}{parsec}
\newacronym{mas}{mas}{milliarcsecond}
\newacronym{nm}{nm}{Nanometer}
\newacronym{CTE}{CTE}{coefficient of thermal expansion}

\newacronym{smc}{SMC}{Small Magellanic Cloud}
\newacronym{lmc}{LMC}{Large Magellanic Cloud}
\newacronym{ism}{ISM}{interstellar medium}
\newacronym{mw}{MW}{Milky Way}
\newacronym{epseri}{$\epsilon$ Eri}{Epsilon Eridani}
\newacronym{EKB}{EKB}{Edgeworth-Kuiper Belt}

\newacronym{CFR}{CFR}{Complete Frequency Redistribution}

\newacronym{nasa}{NASA}{National Aeronautics and Space Agency}
\newacronym{esa}{ESA}{European Space Agency}
\newacronym{omi}{OMI}{\textit{Optical Mechanics Inc.}}
\newacronym{gsfc}{GSFC}{\gls{nasa} Goddard Space Flight Center}
\newacronym{stsci}{STScI}{Space Telescope Science Institute}
\newacronym{nsroc}{NSROC}{\gls{nasa} Sounding Rocket Operations Contract}
\newacronym{wff}{WFF}{\gls{nasa} Wallops Flight Facility}
\newacronym{wsmr}{WSMR}{White Sands Missile Range}

\newacronym{irac}{IRAC}{Infrared Array Camera}
\newacronym[plural=CCDs, firstplural=charge-coupled devices (CCDs)]{ccd}{CCD}{charge-coupled device}
\newacronym{DM}{DM}{Deformable Mirror}
\newacronym{MCP}{MCP}{ Microchannel Plate }
\newacronym{ipc}{IPC}{Image Proportional Counter}
\newacronym{cots}{COTS}{Commercial Off-The-Shelf}
\newacronym{ISR}{ISR}{Incoherent Scatter Radar }
\newacronym{atcamera}{AT}{Angle Tracker}
\newacronym{MEMS}{MEMS}{microelectromechanical systems}
\newacronym{QE}{QE}{quantum efficiency}
\newacronym{RTD}{RTD}{Resistance Temperature Detector}
\newacronym{PID}{PID}{Proportional-Integral-Derivative}
\newacronym{PRNU}{PRNU}{photo response non-uniformity}
\newacronym{DSNU}{PRNU}{dark signal non-uniformity}
\newacronym{CMOS}{CMOS}{complementary metal–oxide–semiconductor}
\newacronym{SIMD}{SIMD}{Single-Instruction, Multiple-Data}
\newacronym{GPGPU}{GPGPU}{General-Purpose computation on Graphics Processor Units}

\newacronym{FOV}{FOV}{field-of-view}
\newacronym{NIR}{NIR}{near-infrared}
\newacronym{PV}{PV}{Peak-to-Valley}
\newacronym{MRF}{MRF}{Magnetorheological finishing}
\newacronym{AO}{AO}{Adaptive Optics}
\newacronym{TTP}{TTP}{tip, tilt, and piston}
\newacronym{FPS}{FPS}{fine pointing system}
\newacronym{SHWFS}{SHWFS}{Shack-Hartmann Wavefront Sensor}
\newacronym{OAP}{OAP}{off-axis parabola}
\newacronym{LGS}{LGS}{laser guide star}

\newacronym{acs}{ACS}{Attitude Control System}
\newacronym{orsa}{ORSA}{Ogive Recovery System Assembly}
\newacronym{gse}{GSE}{Ground Station Equipment}
\newacronym{FSM}{FSM}{Fast Steering Mirror}

\newacronym{WFS}{WFS}{wavefront sensor}
\newacronym{LSI}{LSI}{Lateral Shearing Interferometer}
\newacronym{VVC}{VVC}{Vector Vortex Coronagraph}
\newacronym{VNC}{VNC}{Visible Nulling Coronagraph}
\newacronym{CGI}{CGI}{Coronagraph Instrument}
\newacronym{IWA}{IWA}{Inner Working Angle}
\newacronym{OWA}{OWA}{Outer Working Angle}
\newacronym{NPZT}{N-PZT}{Nuller piezoelectric transducer}
\newacronym{OPD}{OPD}{Optical Path Difference}
\newacronym{WFCS}{WFCS}{Wavefront Control System}
\newacronym{ZWFS}{ZWFS}{Zernike wavefront sensor}
\newacronym{SPC}{SPC}{Shaped Pupil Coronagraph}
\newacronym{HLC}{HLC}{Hybrid-Lyot Coronagraph}

\newacronym{HST}{HST}{ Hubble Space Telescope}
\newacronym{GPS}{GPS}{Global Positioning System}
\newacronym{ISS}{ISS}{International Space Station}
\newacronym[description=Advanced CCD Imaging Spectrometer]{acis}{ACIS}{Advanced \gls{ccd} Imaging Spectrometer}
\newacronym{stis}{STIS}{\textit{Space Telescope Imaging Spectrograph}}
\newacronym{mcp}{MCP}{Microchannel Plate}
\newacronym{jwst}{JWST}{$\textit{James Webb Space Telescope}$}
\newacronym{fuse}{FUSE}{$\textit{FUSE}$}
\newacronym{galex}{GALEX}{$\textit{Galaxy Evolution Explorer}$}
\newacronym{spitzer}{Spitzer}{$\textit{Spitzer Space Telescope}$}
\newacronym{mips}{MIPS}{Multiband Imaging Photometer for \gls{spitzer}}
\newacronym{gissmo}{GISSMO}{Gas Ionization Solar Spectral Monitor}
\newacronym{iue}{IUE}{International Ultraviolet Explorer}
\newacronym{spinr}{SPINR}{$\textit{Spectrograph for Photometric Imaging with Numeric Reconstruction}$}
\newacronym{imager}{IMAGER}{$\textit{Interstellar Medium Absorption Gradient Experiment Rocket}$}
\newacronym{TPF-C}{TPF-C}{Terrestrial Planet Finder Coronagraph}
\newacronym{RAIDS}{RAIDS}{Atmospheric and Ionospheric Detection System }
\newacronym{mama}{MAMA}{Multi-Anode Microchannel Array}
\newacronym{ATLAST}{ATLAST}{Advanced Technology Large Aperture Space Telescope}
\newacronym{PICTURE}{PICTURE}{Planet Imaging Concept Testbed Using a Rocket Experiment}
\newacronym{LITES}{LITES}{Limb-imaging Ionospheric and Thermospheric
Extreme-ultraviolet Spectrograph}
\newacronym{LBT}{LBT}{Large Binocular Telescope}
\newacronym{LBTI}{LBTI}{Large Binocular Telescope Interferometer}
\newacronym{KIN}{KIN}{Keck Interferometer Nuller}
\newacronym{SHARPI}{SHARPI}{Solar High-Angular Resolution Photometric Imager}
\newacronym{IRAS}{IRAS}{Infrared Astronomical Satellite}
\newacronym{HARPS}{HARPS}{High Accuracy Radial velocity Planetary}
\newacronym{hstSTIS}{STIS}{Space Telescope Imaging Spectrograph}
\newacronym{spitzerIRAC}{IRAC}{Infrared Array Camera}
\newacronym{spitzerMIPS}{MIPS}{Multiband Imaging Photometer for Spitzer}
\newacronym{spitzerIRS}{IRS}{Infrared Spectrograph}
\newacronym{CHARA}{CHARA}{Center for High Angular Resolution Astronomy}
\newacronym{wfirst-afta}{WFIRST-AFTA}{Wide-Field InfrarRed Survey
Telescope-Astrophysics Focused Telescope Assets}
\newacronym{GPI}{GPI}{Gemini Planet Imager}
\newacronym{WFIRST}{WFIRST}{Wide-Field InfrarRed Survey Telescope}
\newacronym{HabEx}{HabEx}{Habitable Exoplanet Imaging Mission}
\newacronym{LUVOIR}{LUVOIR}{Large UV/Optical/Infrared Surveyor}
\newacronym{FGS}{FGS}{Fine Guidance Sensor}
\newacronym{STIS}{STIS}{Space Telescope Imaging Spectrograph}
\newacronym{MGHPCC}{MGHPCC}{Massachusetts Green High Performance
Computing Center}
\newacronym{WISE}{WISE}{Wide-field Infrared Survey Explorer}
\newacronym{ALMA}{ALMA}{Atacama Large Millimeter Array}

\newacronym{AURIC}{AURIC}{The Atmospheric Ultraviolet Radiance Integrated Code} 
\newacronym{FFT}{FFT}{Fast Fourier Transform  }
\newacronym{MODTRAN}{MODTRAN   }{ MODerate resolution atmospheric TRANsmission }
\newacronym{idl}{IDL}{$\textit {Interactive Data Language}$}
\newacronym[sort=NED,description=NASA/IPAC Extragalactic Database]{ned}{NED}{\gls{nasa}/\gls{ipac} Extragalactic Database}
\newacronym{iraf}{IRAF}{Image Reduction and Analysis Facility}
\newacronym{wcs}{WCS}{World Coordinate System}
\newacronym{pegase}{PEGASE}{$\textit{Projet d'Etude des GAlaxies par Synthese Evolutive}$}
\newacronym{dirty}{DIRTY}{$\textit{DustI Radiative Transfer, Yeah!}$}
\newacronym{CUDA}{CUDA}{Compute Unified Device Architecture}

\newacronym{MSIS}{MSIS}{Mass Spectrometer Incoherent Scatter Radar}
\newacronym{nmf2}{$N_m$}{F2-Region Peak density}
\newacronym{hmf2}{$h_m$}{F2-Region Peak height}
\newacronym{H}{$H$}{F2-Region Scale Height}

\newacronym{isr}{ISR}{Incoherent Scatter Radar}
\newacronym[description=TLA Within Another Acronym]{twaa}{TWAA}{\gls{tla} Within Another Acronym}
\newacronym[plural=SNe, firstplural=Supernovae (SNe)]{sn}{SN}{Supernova}
\newacronym{EUV}{EUV}{Extreme-Ultraviolet }
\newacronym{EUVS}{EUVS}{\gls{EUV} Spectrograph}
\newacronym{F2}{F2}{Ionospheric Chapman F Layer }
\newacronym{F10.7}{F10.7}{ 10.7 cm radio flux [10$^{-22}$ W m$^{-2}$ Hz$^{-1}$]  }
\newacronym{FUV}{FUV}{ Far-Ultraviolet }
\newacronym{IR}{IR}{Infrared}
\newacronym{MUV}{MUV}{Mid-Ultraviolet }
\newacronym{NUV}{NUV}{Near-Ultraviolet }
\newacronym{O$^+$}{O$^+$}{Singly Ionized Oxygen  Atom }
\newacronym{OI}{OI}{Neutral Atomic Oxygen Spectroscopic State }
\newacronym{OII}{OII}{Singly Ionized Atomic Oxygen Spectroscopic State }
\newacronym{PSF}{PSF}{Point Spread Function}
\newacronym{$R_E$}{$R_E$}{ Earth Radii [$\approx$ 6400 km]  }
\newacronym{RV}{RV}{Radial Velocity}
\newacronym{UV}{UV}{Ultraviolet }
\newacronym{WFE}{WFE}{Wavefront Error}
\newacronym{sed}{SED}{Spectral Energy Distribution}
\newacronym{nir}{NIR}{near-infrared}
\newacronym{mir}{MIR}{mid-infrared}
\newacronym{ir}{IR}{infrared}
\newacronym{uv}{UV}{ultraviolet}
\newacronym[plural=PAHs, firstplural=Polycyclic Aromatic Hydrocarbons (PAHs)]{pah}{PAH}{Polycyclic Aromatic Hydrocarbon}
\newacronym{obsid}{OBSID}{Observation Identification}
\newacronym{SZA}{SZA}{Solar Zenith Angle}
\newacronym{TLE}{TLE}{Two Line Element set}
\newacronym{DOF}{DOF}{degrees-of-freedom}
\newacronym{PZT}{PZT}{lead zirconate titanate}

\newacronym{PCA}{PCA}{Principal Component Analysis}
\newacronym{fwhm}{FWHM}{Full-Width-Half Maximum}
\newacronym{RMS}{RMS}{root mean squared}
\newacronym{RMSE}{RMSE}{root mean squared error}
\newacronym{MCMC}{MCMC}{Marcov chain Monte Carlo}
\newacronym{DIT}{DIT}{Discrete Inverse Theory}
\newacronym{SNR}{SNR}{signal-to-noise ratio}
\newacronym{PSD}{PSD}{Power Spectral Density}

\newacronym{gpu}{GPU}{graphical processing unit}
\newacronym{cpu}{CPU}{central processing unit}
\newacronym{gil}{GIL}{global interpreter lock}

\begin{document} 
\maketitle
\begin{abstract}
Accurately predicting the performance of coronagraphs and tolerancing optical surfaces for high-contrast imaging requires a detailed accounting of  diffraction effects. Unlike simple Fraunhofer diffraction modeling, near and far-field diffraction effects, such as the Talbot effect, are captured by plane-to-plane propagation using Fresnel and angular spectrum propagation. This approach requires a sequence of computationally intensive Fourier transforms and quadratic phase functions, which limit the design and aberration sensitivity parameter space which can be explored at high-fidelity in the course of coronagraph design.  This study presents the results of optimizing the multi-surface propagation module of the open source Physical Optics Propagation in PYthon (POPPY) package. This optimization was performed by implementing and benchmarking Fourier transforms and array operations on graphics processing units, as well as optimizing multithreaded numerical calculations using the NumExpr python library where appropriate, to speed the end-to-end simulation of observatory and coronagraph optical systems. Using realistic systems, this study demonstrates a greater than five-fold decrease in wall-clock runtime over POPPY's previous implementation and describes opportunities for further improvements in diffraction modeling performance. 
\end{abstract}

\section{INTRODUCTION}
\label{sec:intro}  

The Physical Optics Propagation in PYthon module (POPPY) \cite{perrin_simulating_2012,perrin_poppy_2016} is an open-source library for modeling optical diffraction. POPPY was originally developed as the back-end diffraction code for the package WebbPSF, which simulates \gls{PSF}s for the James Webb Space Telescope (JWST) and now also the Wide Field Infrared Survey Telescope (WFIRST), but its simulation capabilities are broadly applicable across many kinds of astronomical simulations and beyond (for instance, there are users of POPPY working on topics such as modeling laser physics and fiber optic coupling). Accurate models of \gls{PSF}s are needed throughout an instrument or observatory's lifetime, supporting tasks such as mission design and instrument trade studies, planning of scientific observations, and data analyses such as PSF fitting photometry and astrometry or deconvolutions. 

Coronagraphic observations place particularly stringent demands on such simulations, due to the inherent complexity of coronagraphic optical systems and the need for very high optical fidelity given ambitious contrast goals such as the 1e-10 needed for future imaging of rocky planets in reflected light. This increased fidelity comes at a cost of correspondingly greater computational cost.  In order to enable efficient design searches over large parameter spaces, this paper describes our progress in substantially reducing the run time of coronagraphic diffraction simulations by incorporating improved state-of-the-art numerical libraries and by porting computationally-intensive portions of the algorithm to run on highly parallel graphics processing units (GPUs). These new accelerated calculations are available now in POPPY release 0.7, with source code available on Github and software distribution through the Conda and PyPI systems. 
POPPY provides a python interface to a range of diffraction models through familiar AstroPy\cite{the_astropy_collaboration_astropy_2013} based scientific Python tools, including definition of optical surfaces  as FITS files and automatic handling of units.

In the following sections we summarize the computational context for modeling optical propagation, describe our methods for accelerating calculations in POPPY, and present performance benchmark results on a few representative cases. 
In particular, we show an example of Fresnel propagation through an optical model representative of the WFIRST Coronagraphic Instrument using a Shaped Pupil Coronagraph. 

\subsection{Implementing Fresnel Propagation Methods}

POPPY supports optical simulations in two regimes, the standard Fraunhofer (far-field) and Fresnel (near-field) approximations \cite{goodman_introduction_2005}. The methods used in Fraunhofer calculations have been previously presented\cite{perrin_simulating_2012}. To quickly generate results in the Fraunhofer regime, for coronagraphs with a small focal plane mask size POPPY typically employs the fast semi-analytic coronagraph propagation algorithm from  Soummer et al\cite{soummer_fast_2007} which uses discrete matrix Fourier transforms of modest sized arrays. Some Fraunhofer regime calculations also use standard \gls{FFT}s of larger arrays, particular in cases where the coronagraph geometry is incompatible with the semi-analytic algorithm (e.g. the four quadrant phase mask coronagraphs used on JWST MIRI).

The highest fidelity coronagraphic simulations require modeling propagation in the Fresnel regime. Fresnel propagation is necessary to account for effects such as mixing between phase and amplitude errors, as described below. Extending POPPY to support Fresnel regime calculations was a straightforward process, taking advantage of Python's flexible object-oriented programming to provide a general complex wavefront class and interfaces to a variety of optical systems.  POPPY now supports Fresnel calculations using both the direct Fresnel and the angular spectrum propagation methods. We implement plane-to-plane  angular spectrum propagation following the methods in Ref.~\cite[Equation 82-89]{lawrence_optical_1992}, under the assumption that powered optics can be treated as thin lenses \cite[Equation 5-10]{goodman_introduction_2005}. An automated software continuous integrations system reproduces a series of textbook examples and other test cases after every code commit to rigorously validate correct operations. 
POPPY's implementation of Fresnel propagation has also been cross-checked against other implementations such as PROPER\cite{krist_proper_2007}. The code is open source under a standard BSD 3-clause license, freely available for academic and commercial uses, and we welcome community participation in its ongoing development\footnote{See \url{https://poppy-optics.readthedocs.io/} for documentation, installation instructions, and source code.}. 

\begin{figure}[htp]
\centering
\begin{subfigure}{\textwidth}
\centering
\includegraphics[width=.7\linewidth]{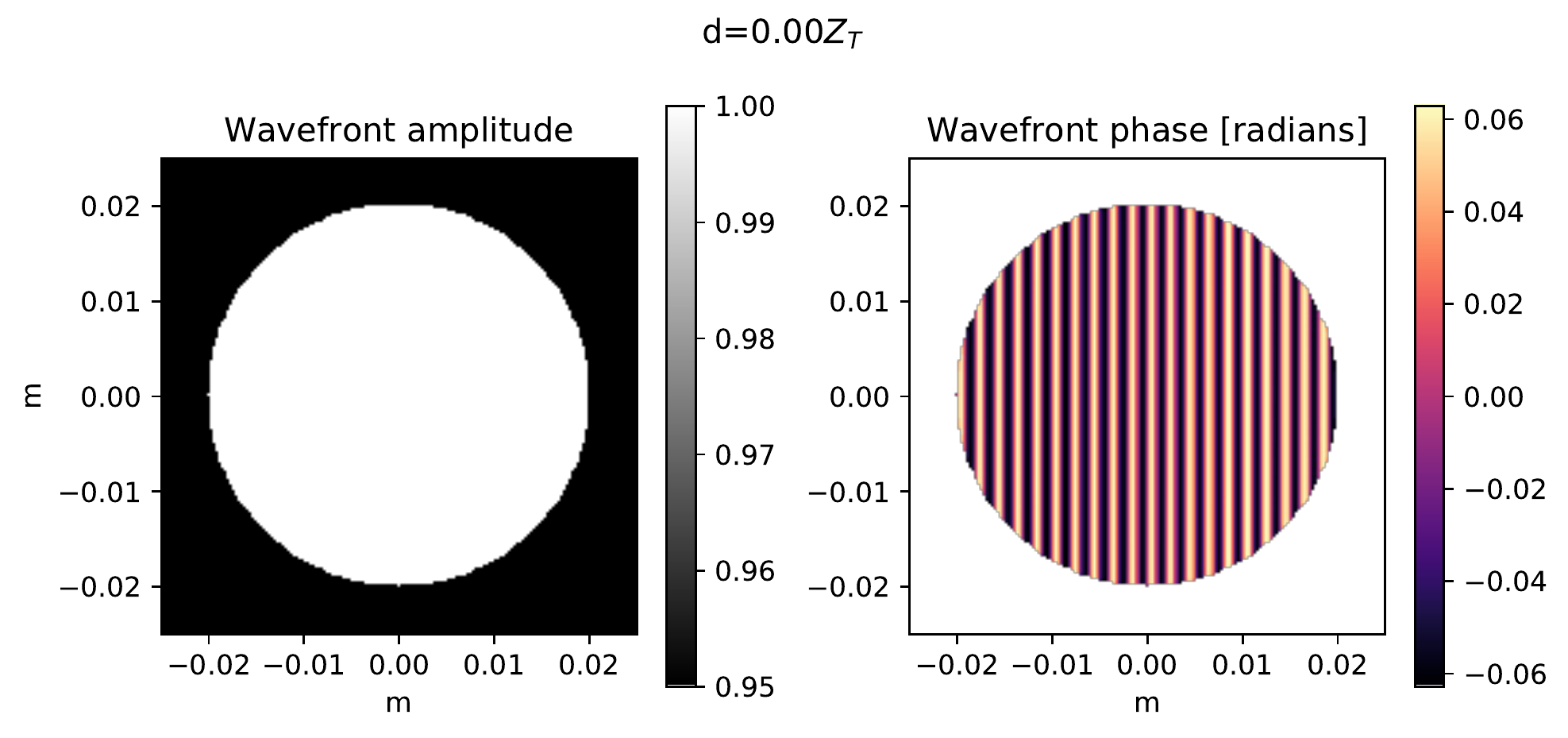}
\end{subfigure}
\begin{subfigure}{\textwidth}
\centering
\includegraphics[width=.7\linewidth]{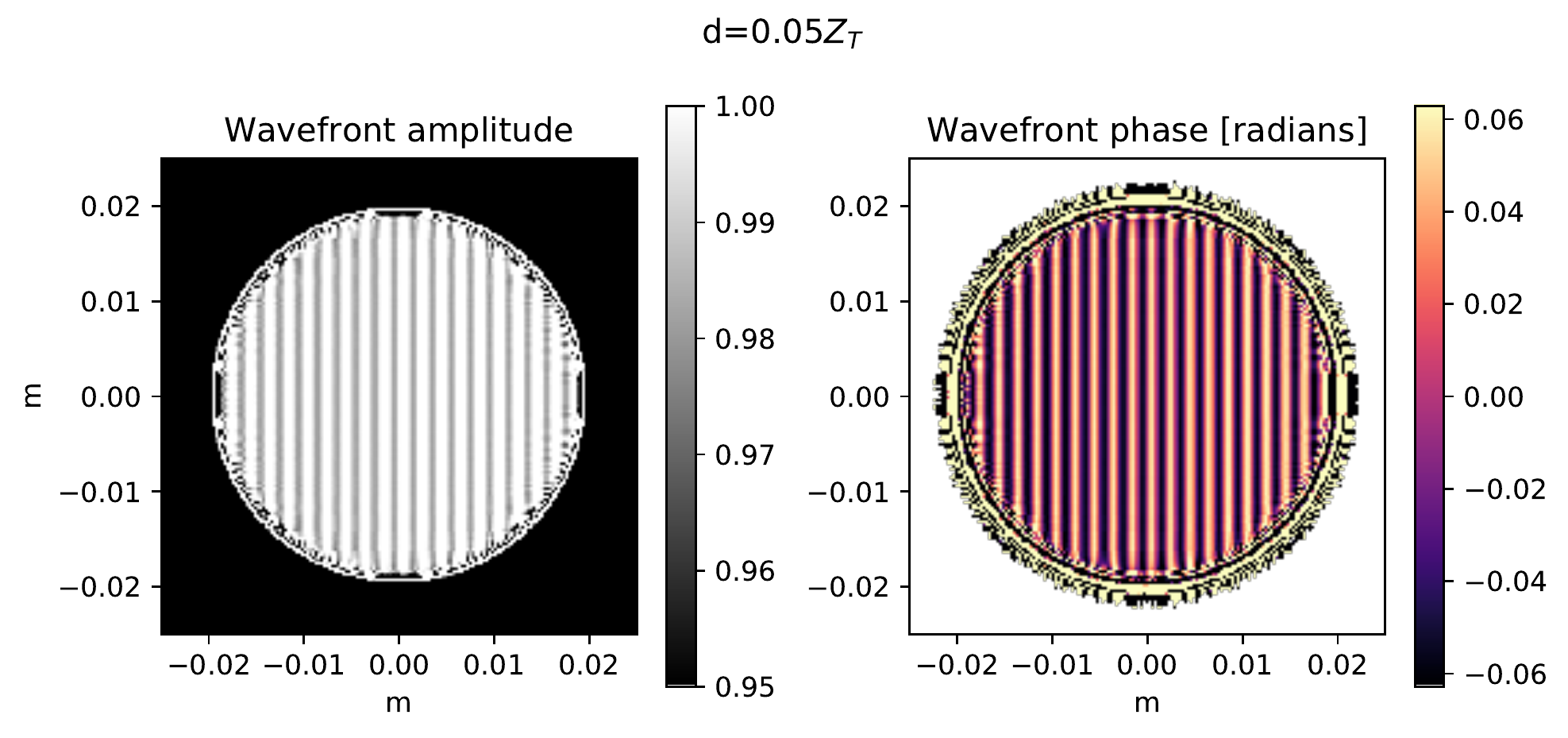}
\end{subfigure}
\begin{subfigure}{\textwidth}
\centering
\includegraphics[width=.7\linewidth]{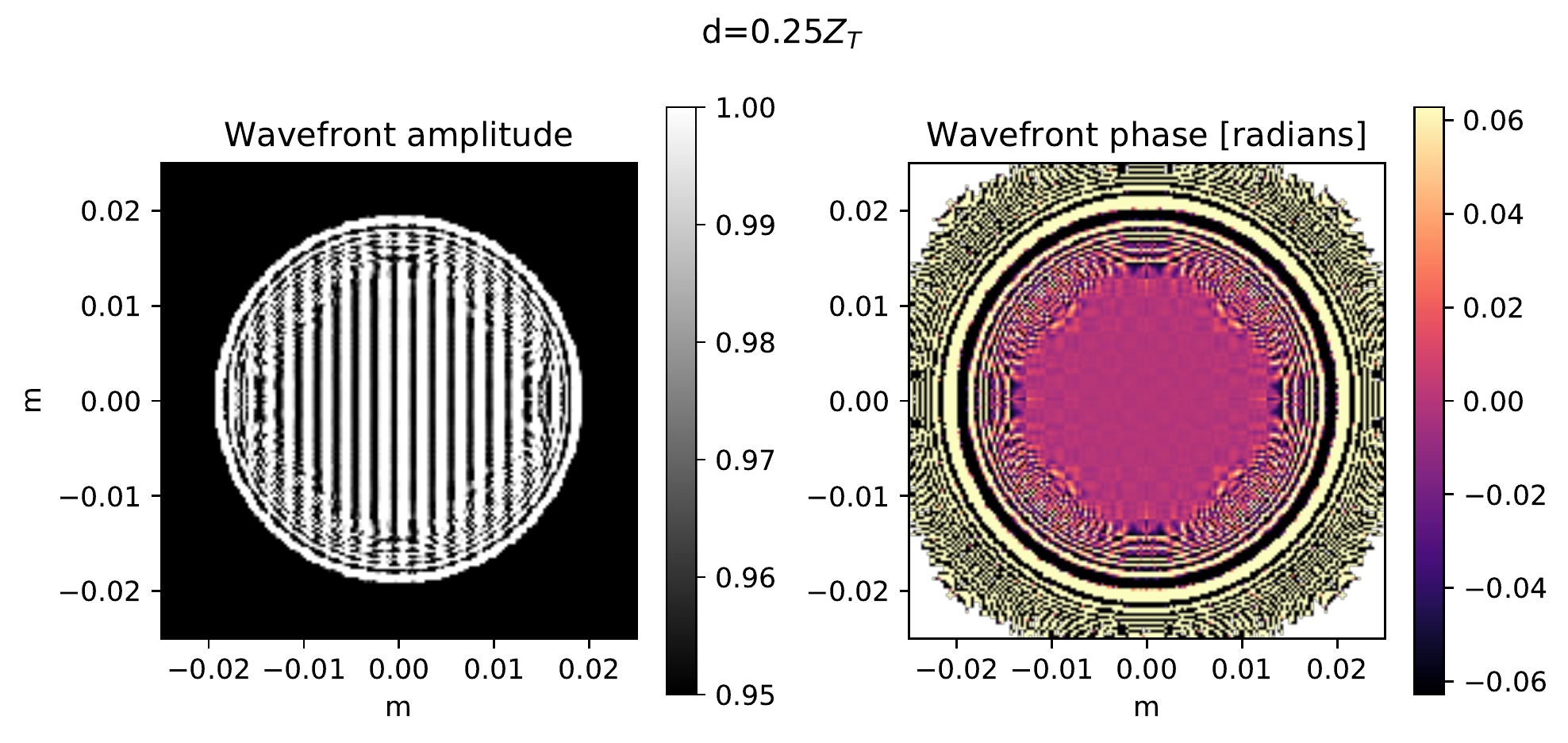}
\end{subfigure}
\caption{Example of the Talbot Effect calculated using the POPPY Fresnel module. 
In addition to edge effects, phase errors in the plane at $Z=0$ (top row) are  partially converted into  a similar pattern of amplitude errors at an intermediate Rayleigh range (middle row) and completely converted at a range of $Z=0.25 Z_T$ (bottom row).
The POPPY prescription used to perform this example calculation is presented in Appendix A.  }\label{fig:talbot}
\end{figure}

Plane-to-plane angular spectrum diffraction methods have become standard in simulating  high-contrast coronagraph designs\cite{marois_end--end_2008,yaitskova_foros_2010,krist_assessing_2011,douglas_end--end_2015,mendillo_optical_2017,lumbres_modeling_2018}.
Unlike Fraunhofer diffraction methods, the angular spectrum method captures the transformation of optical phase error into amplitude errors, the ``Talbot effect". 
As described elsewhere, e.g. \cite[Eq. 1]{ndiaye_high-contrast_2013}, a characteristic Talbot length is given by
 \begin{equation}
Z_T =2(D/f)^2/\lambda
 \end{equation}
 Where D is the aperture diameter, $f$ is the spatial frequency and $\lambda$ is the wavelength.
Periodic wavefront errors function as a diffraction grating and at propagation distance $d=Z_T$ from the input,  phase errors will completely transform into amplitude errors. 
To illustrate the Talbot effect using POPPY, we setup a system with a 2 cm aperture and a sinusoidal wavefront error with an amplitude of 5 nanometers and 20 cycles per aperture, approximating one dimension of actuator print through or ``scalloping'' of a \gls{MEMS} deformable mirror. The result is illustrated in Fig. \ref{fig:talbot}: the top panel shows the input amplitude and phase maps, the middle panel shows the amplitude and phase after propagating $0.05Z_T$ and the bottom panel shows complete conversion from phase to amplitude error, at 0.25$Z_T$ as expected.
Additionally, diffraction from the edge of the circular aperture becomes more pronounced at the larger distance and is seen as circularly symmetric ripples. These effects must be taken into account when modeling high performance coronagraphs.

\subsection{Computation Needs for Modeling Optical Propagation}

Such calculations are dominated primarily by a sequence of Fourier transforms of large 2D arrays of complex floating point numbers. Additional significant computation costs come from evaluations of complex exponential functions, and the general array operations for creating and manipulating the arrays representing coronagraph masks and wavefronts.

For high-contrast imaging and wavefront and control using deformable mirrors, it is necessary to sample each optic to spatial frequencies significantly higher than the deformable mirror actuator spacing. 
  Space coronagraph missions under development such as WFIRST and PICTURE-B/C\cite{noecker_coronagraph_2016,mendillo_optical_2017,douglas_wavefront_2018} employ deformable mirrors with 32$\times$32 or 48$\times$48 actuators and sample the pupil with hundreds of pixels\cite{krist_technology_2013} across the aperture, add padding as appropriate to prevent aliasing and allow $N\times N$ arrays with $N$ of typically 2048 or 4096.
Following ground-based coronagraphs to higher actuator counts\cite{macintosh_gemini_2008} later space missions with larger apertures may employ deformable mirrors with actuator per axis of 64 to upwards of 196 \cite{morgan_initial_2015, bolcar_large_2017}. 
Accurately sampling these spatial frequencies will require significantly larger arrays. 
For example, for a 196$\times$196 actuator to achieve the same seven pixel-per-actuator sampling used in past plane-to-plane simulation of \gls{VVC} coronagraphs\cite{krist_technology_2013} and maintain a base-2 array dimensions would require  $N=8192$.

The well-known Cooley-Tukey \gls{FFT} algorithm\cite{cooley_1965} enables computing discrete Fourier transforms with $O(n \log n)$ operations, as opposed to $O(n^2)$ for direct evaluation of the discrete Fourier transformation, where $n$ is the total number of elements in the array being transformed (i.e. for 2D arrays we have $n=N\times N$, the product of the array dimensions). The computational savings from the \gls{FFT} algorithm are immense, saving a factor of about a million for $N=4096$ compared to the plain DFT. But still, each doubling of the length per axis $N$ results in $\sim 4-5\times$ more CPU operations per transform. In addition to processing time, such calculations require significant memory.

Typically many such transformations are necessary. For instance, the WFIRST CGI example propagation presented below, with a total of 10 optics, requires a total of 19 \gls{FFT} for the propagation. Further, this must then be repeated for all of the wavelengths simulated. 

Sampling of optical elements is also critical. 
As illustrated by Fig \ref{fig:talbot}, wavefront errors on optics are commonly the motivation for plane-to-plane simulations.
POPPY will interpolate input optical element surface maps to match the incident wavefront sampling, but such interpolation is inefficient and currently relies on single-core SciPy routines\cite{jones_scipy_2001}.
We suggest determining the wavefront sampling which well captures the spatial frequencies of interest, and the spatial sampling of the POPPY wavefront incident on each optic before generating matching wavefront error maps. 
This can be done empirically by running a trial system in POPPY and inspecting the
\textit{pixelscale} attribute  of the intermediate wavefront at each optic.

\subsection{Achieving High Performance in Scientific Python}

As the Python language has become widespread in scientific, engineering, and educational computing\cite{fangohr_comparison_2004,greenfield_reaching_2007,hirst_ureka_2014-1}, there have been widespread efforts towards enabling high performance numerical work. In fact astronomers played a significant role\cite{Greenfield_2003,greenfield_reaching_2007} in the early development and establishment of the \texttt{numpy} library, which has now become the de facto standard for working with numerical arrays in Python, along with the broader SciPy ecosystem\cite{jones_scipy_2001}.
Python as a high-level interpreted language offers tremendous practical advantages in speed of development and flexibility for interactive use, and with the proper methods it is possible to achieve speed rivaling that of code written in compiled languages such as C or Fortran. Often this is done by linking to compiled code to implement low-level array operations.  NumPy and Scipy provide interface to C libraries for efficient calculations, including options to link against highly optimized mathematical libraries such as the Intel Math Kernel Library, as discussed below. Linking against a well-optimized linear algebra library can provide order-of-magnitude gains in performance. 

Broadly speaking, achieving good performance generally requires carefully understanding both the details of the computing hardware and of the workload. For instance some tasks are rate-limited by CPU computation, versus others by data input/output (IO). To accelerate CPU-limited tasks we want to efficiently use features of the hardware such as multiple processors, vectorized operations (e.g. SSE and AVX in recent Intel CPUs), or the larger suite of registers available in 64-bit architectures.  Efficiently using modern hardware requires workloads that can be parallelized across multiple compute cores, for both CPU and GPU architectures. To accelerate IO-limited tasks, we want efficient memory access, for instance making good use of fast low-level caches. Sometimes these goals are at odds; for instance there is an overall cost in IO to transfer and synchronize data between multiple processors. 

The most commonly used implementation of Python, CPython, employs a so-called \gls{gil}\cite{_globalinterpreterlock_2018}. 
The \gls{gil} prevents multiple threads from simultaneously executing commands, effectively preventing CPython code from executing using more than one processor at a time. This can be worked around by running multiple Python processes in parallel via the \texttt{multiprocessing} module. POPPY does support this method for parallelizing multiwavelength calculations, but it incurs additional overhead from process startup and interprocess communications. Running multiple threads generally has better performance than multiple processes; for Python, this is done by linking against multi-threaded compiled code that avoids the GIL limitation.

In order to better leverage multiple processors while preserving the convenience and readability of Python, a variety of libraries for high performance computing have been developed including NumExpr\cite{cooke_numexpr_2018}, and  Numba\cite{lam_numba_2015} a just-in-time compiler that can target either \gls{cpu}s or \gls{gpu}s. 
NumExpr accelerates many element-wise array calculations even more than NumPy, by breaking large arrays into smaller chunks and parallelizing the mathematical operation across multiple threads. The chunk sizes are selected to efficiently make use of CPU cache memory, which both helps accelerate calculations and minimizes the need to allocate additional memory for intermediate results. 

Recently, \gls{gpu}s have made significant inroads into scientific computing, providing many parallel processor cores on a single board. For many workloads, speed ups of a factor of a few times to perhaps $10\times$ can be achieved. Introduced in 2007, the NVIDIA \gls{CUDA} library\cite{nickolls_scalable_2008} provides a C-like interface to parallelized operations distributed across hundreds to thousands of cores.
OpenCL\cite{stone_opencl_2010} provides a  portable framework writing code to run on a variety of architectures, including \gls{cpu}s, \gls{gpu}s and Field Programmable Gate Arrays (FPGAs). \gls{CUDA} is the leading architecture for \gls{GPGPU} efforts curently, but requires NVIDIA hardware. OpenCL works on a wider range of hardware, in particular the AMD GPUs available in typical Apple computers which are often used by many astronomers.

Previous authors have implemented diffraction simulations directly in \gls{CUDA}\cite{shimobaba_numerical_2008}. We have combined the user friendly POPPY optical system and wavefront classes with optional high performance computing libraries, to provide a framework which can easily be developed on a laptop computer and builds on and accelerates existing diffraction modeling libraries such as WebbPSF\cite{perrin_simulating_2012}.

The Intel Math Kernel Library \cite{pavlyk_accelerating_2017}, particularly the multi-core Intel \gls{FFT} libraries, can also provide significant speed ups, equalling or exceeding FFTW in certain cases. This is particularly true for recent high-end processors which include 512-bit-wide vector arithmetic units for fast \gls{SIMD} arithmetic operations.  Which method is optimal in a given case will depend on both the hardware available and the size of the arrays involved.

\section{Methods}

\subsection{Profiling Enables Prioritization of Optimization Efforts}
Efficiently targeting optimization work requires an accurate understanding of which parts of program dominate the costs.\footnote{Hence the famous aphorism attributed to Donald Knuth: ``Premature optimization is the root of all evil''.} In order to decrease the overall run-time of POPPY  calculations we therefore ran a set of test cases using IPython's  \texttt{\%prun} profiling ``magic'' command which calls Python's built-in code profiler to measure the number of function calls and the total time per call during code execution. This profiling resulted in the identification of three major bottlenecks: 
\begin{enumerate}[itemsep=-1ex]
\item Calculations of the discrete Fast Fourier Transform,
\item Calculations of complex exponentials and related functions, 
\item Calls to the \texttt{fftshift} functions to reorder arrays.
\end{enumerate}

\noindent We therefore set out to optimize each of these tasks. 

\subsection{Accelerating the Discrete Fast Fourier Transform: GPUs and/or Intel MKL}
The first area to optimize is the calculation of 2D Fourier transforms of large complex arrays. 
The default \gls{FFT} option provided with NumPy is implemented via an interface to the FFTPACK C library. 
A generally faster library, FFTW \cite{frigo_design_2005} offers parallelized computation of the discrete Fourier transform in $O(N\log N)$ for any  $N$ (not limited to powers of two). FFTW can be called from Python using the \texttt{pyFFTW} package (https://pypi.org/project/pyFFTW/). FFTW is widely regarded as a state-of-the-art \gls{FFT} library for execution on CPUs, and provides highly tuned multi-threaded FFTs and automatic capabilities for tuning algorithm parameters to optimize performance on individual compute devices. POPPY has long supported FFTW, falling back to NumPy's FFT library when FFTW was unavailable. 

In order to further accelerate calculation of the discrete Fourier transform, we implemented the use of FFTs on GPUs via first \gls{CUDA} and then also OpenCL. Subsequently and very recently, we identified an alternative pathway using Intel's Math Kernel Library.

\noindent \textbf{Fast FFTs on GPUs with CUDA and OpenCL:}
FFTs on GPUs  typically provide up to an order of magnitude advantage over FFTW\cite{steinbach_gearshifft_2017}, particularly if high-end NVIDIA GPUs are used. 
The \texttt{pyculib} library\cite{_pyculib_2018} (formerly Anaconda Accelerate\cite{_anaconda_}) provides a python wrapper around the NVIDIA cuFFT Library\cite{_cufft_2018}, allowing parallel computation of FFTs on a GPU. Similarly, the \texttt{pyopencl} and \texttt{gpyfft} packages together provide an interface to the OpenCL library clFFT\cite{_clfft_2018}.
Interfaces to both of these have been implemented in POPPY, in a way that is transparent to users; calculations will take advantage of a GPU if available, without requiring any direct changes in the user's optical system model code.  
Benchmarks for these are presented below. 

An important factor to consider is the large variations in capabilities between different GPUs, and different classes of GPUs. Some devices, such as NVIDIA's Tesla K80, are specifically designed for high performance, high throughput scientific and industrial computing, with very high core counts and substantial resources for double-precision (64-bit) floating point arithmetic. 
These will provide the highest performance in cost-insensitive environments and for limited duration calculations are now widely available for rent from many cloud computing services.
 Other GPUs place lower priority on double-precision performance, including some devices which are relatively high-end but which primarily target single-precision (32-bit) performance for 3D graphics and other media authoring workloads. For instance, the Radeon Pro Vega 56 (one of AMD's current top-end cards, currently used in Apple's iMac Pro workstations) has only a subset of its processing cores that are capable of double-precision math. Our testing of FFTs on this device shows a drop of 3-5x in speed for double vs single precision floats, consistent with previous published benchmarks by others. (Not all optical calculations require double precision math, and POPPY does have support for single precision calculations if desired. But highly demanding space coronagraphy simulations merit the use of double-precision floating point. We therefore focus only on that mode in this work.)

\noindent \textbf{Fast FFTs on CPUs with Intel MKL:} Intel has long developed its own highly optimized library of mathematical functions, to Intel Math Kernel Library (MKL). Until recently, licensing restrictions prevented it from becoming a widely used standard in Python. However that is no longer the case, and Intel's optimized code is now freely available via the Anaconda Python package distribution system, as well as from Intel's own distribution channels\cite{_intel_-1}.
\footnote{That said, ensuring that you have installed and are using a version of numpy that uses MKL can still be somewhat opaque and confusing. Users should observe the number of cores in use during FFT operations and check that conda has installed both the \texttt{mkl} and \texttt{mkl\_fft} packages, and that \texttt{numpy.\_\_config\_\_.show()} reports that \texttt{numpy} is indeed linked against MKL.
}

Intel MKL includes a very highly tuned FFT implementation, which takes full advantage of the full set of features in modern processors including multithreading, highly vectorized \gls{SIMD} operations, and efficient memory access and caching. NumPy can be compiled to make use of this FFT implementation directly to replace its prior FFTPACK-based FFT implementation, avoiding any need to change code. Somewhat to our surprise, we have found that using the most recent builds of NumPy 1.14, Python 3.6, and MKL 2018.0.2, FFT speeds directly in NumPy can surpass those achieved with FFTW, and in some cases even surpass GPUs. Intel MKL now appears easy to recommend as a high performance solution for many users, given its relative ease of installation and much simplified development compared to writing code for GPUs.  But for the largest calculation sizes, it cannot match the highly parallelized performance of high-end GPUs.

\subsection{Accelerating Evaluations of Exponential Expressions: NumExpr}

After optimizing FFT runtime, the next most time-consuming operation is evaluations of exponential functions. 
This occurs both as the exponential term in the quadratic phase factor,
\begin{equation}
\exp[i(x^2 + y^2)/z],
 \end{equation}
  where $x$, $y$ and are $N\times N$ arrays and $z$ is a real number, and also in evaluating the complex phasor of a given optical element,
\begin{equation}
A \exp[2 \pi i / \lambda \cdot \phi ],
 \end{equation}  
where $A$ and $\phi$ are again  $N\times N$ arrays and $\lambda$ is a real number.

We compared run times for $N$=2048 using NumPy, NumExpr, and Numba for both processor and GPU (\url{http://numba.pydata.org}). 
The NumExpr library provided the largest speed gain (followed closely by a \gls{CUDA} compiled Numba call) on our test machine. As stated above, NumExpr works by evaluating such expressions in chunks, enabling both efficient use of high-speed cache memory and multi-threaded operations. The HOPE package\cite{akeret_hope_2015}, a specialized just-in-time compiler developed at the ETH Zurich Institute of Astronomy, provided faster evaluation of this function under Python 2.7, but it is no longer actively being developed and does not currently work on Python 3.6.

POPPY was therefore adapted to use NumExpr for such calculations. Switching to NumExpr results in gains of $4-10\times$ in speed compared to numpy for evaluating such expressions. Machines with more processor cores typically see greater speedups, with diminishing returns above perhaps 12 to 16 cores depending on the size of the array. 

Given how well NumExpr worked to accelerate this part of the calculation, and its very straightforward ease-of-use, we also adopted NumExpr in several other parts of POPPY. For instance calculating the radius is necessary for generating typical circular optics, while generating large arrays of sines and cosines are needed in the matrix DFT used in the Fraunhofer code. Even simple products of multiple arrays, or of an array and a scalar, can often be accelerated by using NumExpr. NumExpr intentionally supports only a strictly limited set of array operations, but for that subset it is very highly performance.

\subsection{Accelerating Fourier Array Shifting on the GPU}
The FFT algorithm imposes certain fixed relationships on the locations of spatial frequencies within the output array, in particular locating the origin in the extreme corners of the array rather than in the center. 
Shifting the array contents to relocate the zero-frequency origin to the center of a Fourier transformed spectrum--the so-called ``fft shift'' operation--allows easier application of 2D transmission and wavefront error maps at each optic, and also enables more intuitive display and interactive examination of intermediate results. In profiling, the numpy \texttt{fftshift} function was found to be a significant contributor to calculation time.

Using a Numba just-in-time decorator to implement the \gls{CUDA}  of the FFT algorithm by Abdellah\cite{abdellah_cufftshift_2014} the time to recenter arrays of size 4096$\times$4096 or 8192$\times$8192 was decreased by a factor of approximately three. The highly multithreaded process on the GPU wins out over the additional IO overhead for transferring the arrays to and from the GPU. 
Note that as of this writing, the \texttt{fftshift} operation on GPU is only supported for CUDA, though extending this to OpenCL as well is planned for the near future.

\subsection{Reproducibility}
The benchmarking results and figures reported here were generated in a Jupyter notebook which the reader is invited to review and test on other system\cite{douglas_douglase/poppy_benchmarking_2018}. 
Benchmarking results were measured using the IPython\footnote{version 6.1.0 in Python 3.6} \textit{\%timeit} magic command on default settings. 
For each configuration a mean and standard deviation is reported for seven runs of each function, with ten loops per run for functions that return in under 0.2 seconds (and additional loops in powers of ten until 0.2 seconds is reached).


\section{Results}

\subsection{Fresnel Optical System Run Times}

\begin{figure}
  \caption{Typical NumPy run times without MKL-FFT, FFTW, NumExpr, or CUDA accelerations for the example coronagraphic system (Appendix B) on an 8 Core 2.0 GHz Intel Sandy Bridge Xeon machine running on the Google Compute Engine. Times given are for the full optical system calculation with many steps, not just an $N\times N$ \gls{FFT} calculation. (Note, multi-core MKL \gls{FFT} were not used for this baseline). }\label{fig:numpyruntime}   
    \begin{subfigure}[h]{0.45\textwidth}
  \includegraphics[width=1\textwidth]{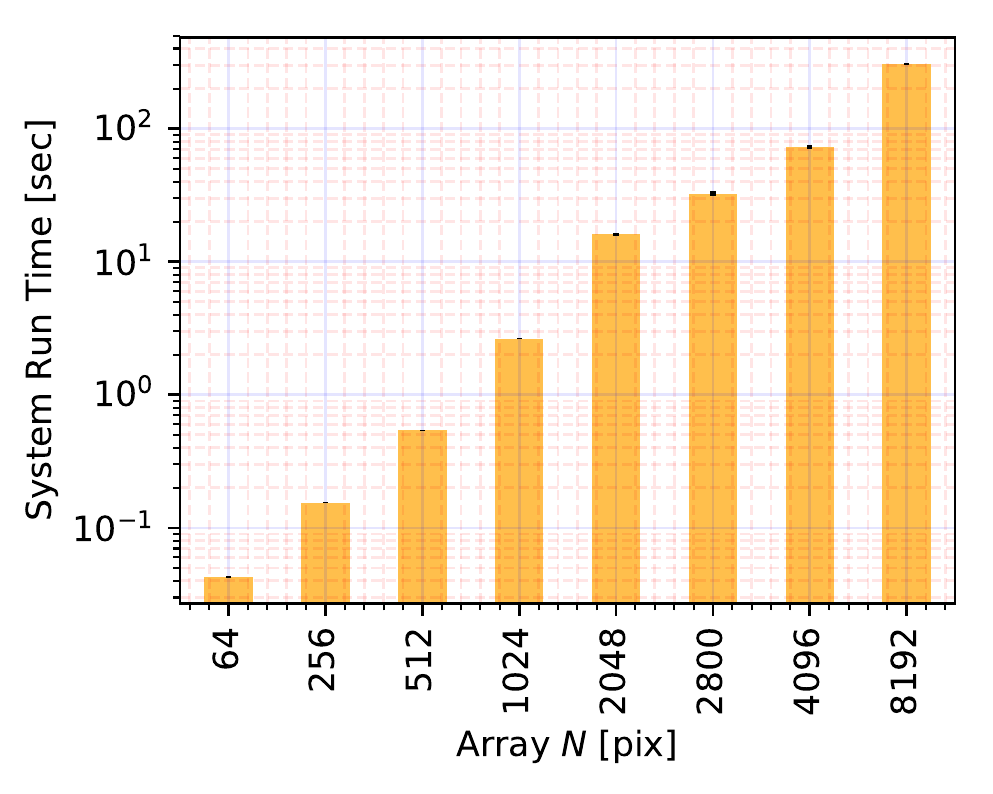}
  \end{subfigure}
      \begin{subfigure}[t]{0.45\textwidth}
      \begin{minipage}[t]{\linewidth}\centering
\begin{tabular}{lrr}
$N$ &       Average [sec] & 1$\sigma$ [sec]\\
\midrule
64   &    0.1069 &  0.001815 \\
256  &    0.2423 &  0.004003 \\
512  &    0.7555 &  0.007441 \\
1024 &    3.3480 &  0.027630 \\
2048 &   19.7900 &  0.251200 \\
2800 &   40.3000 &  0.210700 \\
4096 &   90.1600 &  0.374700 \\
8192 &  376.0000 &  2.576000 \\
\bottomrule
\end{tabular}
\end{minipage}
  \end{subfigure}
  
  \end{figure}

Measured mean \textit{unoptimized} run times for the Fresnel coronagraphic test system, Appendix \ref{code:fresnel}, are plotted and tabulated in Fig. \ref{fig:numpyruntime}. 
The runtime increases approximately as the area of the array ($\propto N^2$).
Notably, the $N=2800$ runs faster than the next power of two, indicating significant speed gains maybe found by not requiring power of two array sizes in some cases.

\begin{figure}
  \caption{Example 10 optic system (Appendix B) run time versus array size, normalized to a standard NumPy installation's run times )Fig. \ref{fig:numpyruntime}).}  \centering
    \begin{subfigure}[b]{0.45\textwidth}
        \includegraphics[width=1\textwidth]{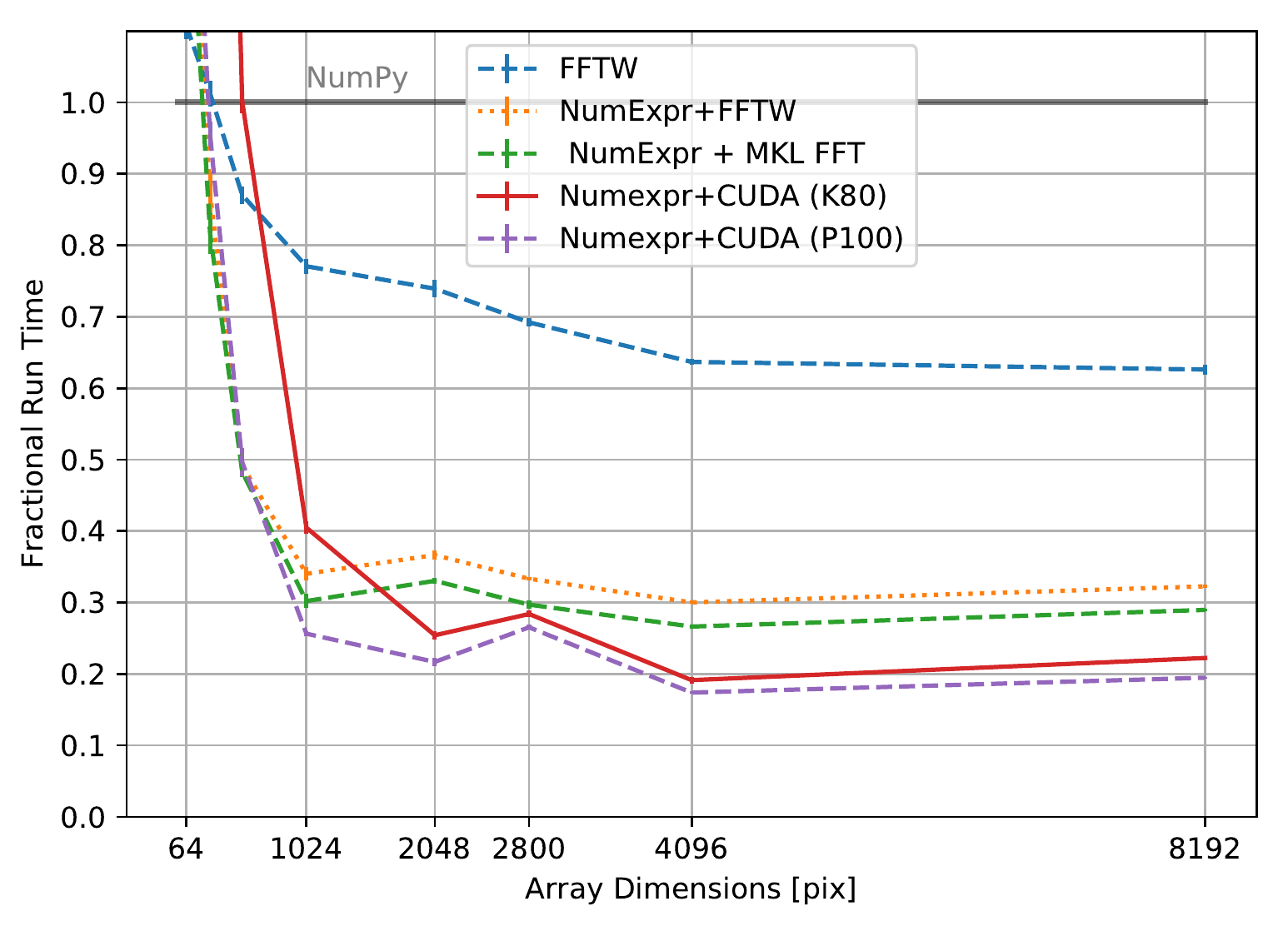}
                \caption{Eight Core Intel Xeon 2.0GHz machine running on the Google Compute Engine. A NVIDIA K80 GPU was used for the for CUDA operations (lower solid curve). The bottom dashed curve shows a small but consistent improvement improvement using a newer, P100 GPU on a 2.2 GHz Xeon Machine with 16 cores. }\label{fig:benchmarksGCE}
                  \end{subfigure}
                  \hfill
     \begin{subfigure}[b]{0.45\textwidth}
         \includegraphics[width=1\textwidth]{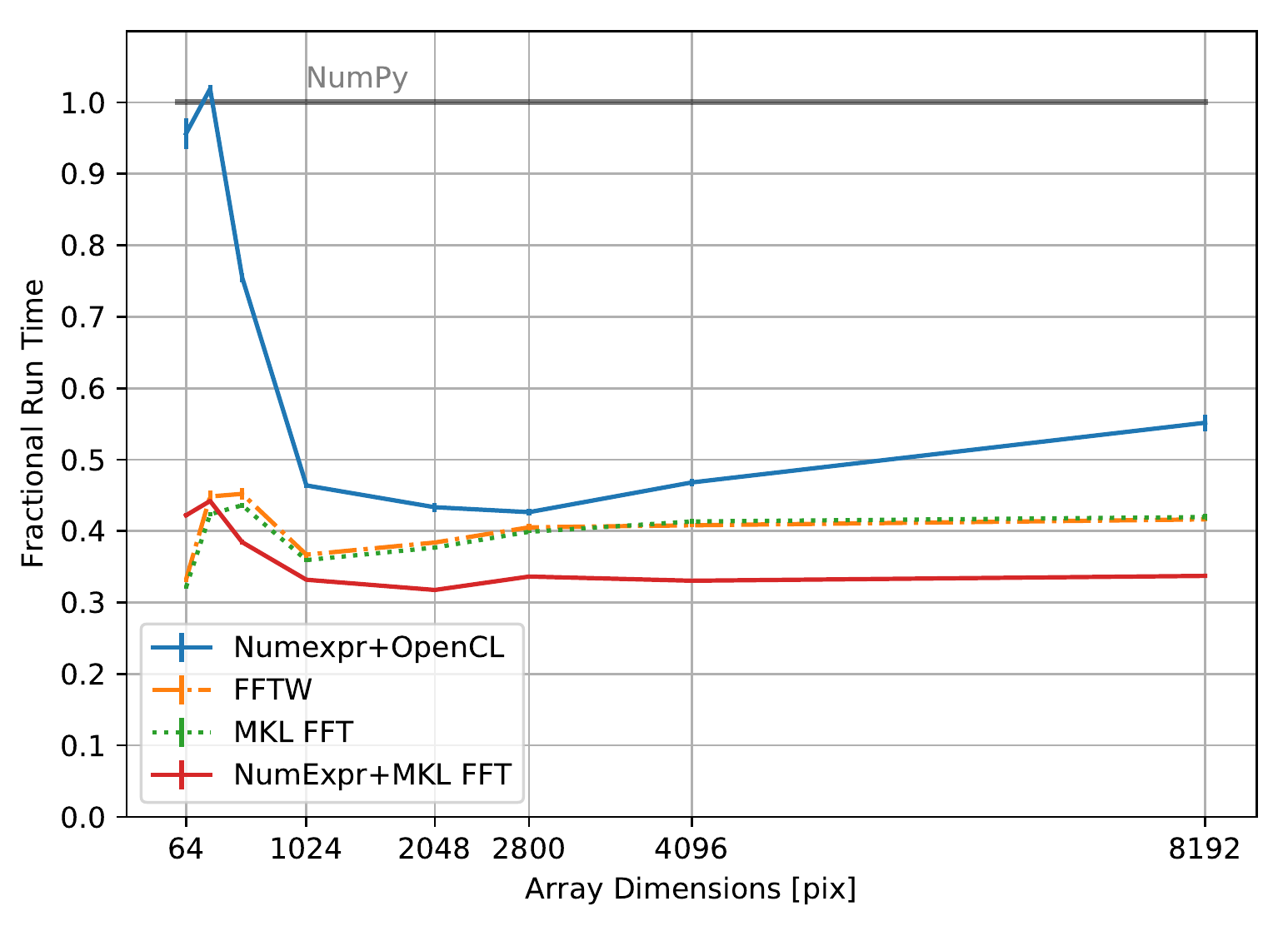}
 \caption{Eight core iMac Pro, Intel Xeon 3.2GHz with AMD RadeonPro Vega 56.
 OpenCL GPU \gls{FFT} performance does not match the speed up from Intel MKL \gls{FFT} on this system and does not include sn optimized \textit{fftshift} function. Arrays smaller than 1024 see a greater speedup than on the 2.0GHz Xeon machine.}\label{fig:MacPro}           \end{subfigure}

\end{figure}

The results of the successive optimizations are shown in Fig. \ref{fig:benchmarksGCE}. 
Each curve is normalized to the NumPy runtime (e.g. Fig. \ref{fig:numpyruntime}) for the same values of $N$.
For the smallest arrays tested switching to FFTW gives no advantage, in fact, additional overhead can an slow the run time. 
However, for large arrays FFTW runs in approximately 75\% of the time required for standard NumPy.
MKL-FFT run times are slightly faster.
In place array calculations were identified as a bottleneck during profiling, and this is illustrated by the significant increase in speed using NumExpr (dot dash orange line), taking less than 40\% of the time for N$>$1024.

GPU runtimes using \textit{pyculib} and the CUDA \textit{fftshift} algorithm described above on a K80 NVIDIA GPU provide an approximately factor of five increase over base NumPy.

Newer generation GPUs can provide additional speed up, at higher purchase or rental prices. 
The bottom-most dashed line in Fig. \ref{fig:benchmarksGCE}, but for a newer NVIDIA P100 GPU on an 16 Core Intel Sandy Bridge 2.2 GHz system.


Fig. \ref{fig:MacPro} shows the performance improvements for an 8-core iMac Pro, Intel Xeon 3.2GHz with AMD RadeonPro Vega 56. 
Small arrays are faster than on the 2.0GHZ    machine but the consumer grade GPU does not provide a performance improvement over CPU calculations.

\section{Conclusions}

In the initial, pre-optimization code state, calculation times were typically dominated by the FFTs (in plain NumPy or FFTW). After the optimizations described in this paper, the FFT has been accelerated enough that the calculations of complex exponentials are often a more substantial contributor. 

 In the era of readily scalable cloud computing resources, optimizing the financial cost of a simulation ensemble is often important. 
In addition to increased efficiency, some of the optimizations, particularly the GPU optimizations scale with system cost. 
For parallelized diffraction modeling operations, such as multi-wavelength calculations, it is important to weigh the relative cost of renting a large number of less expensive machines against the performance gains of more expensive systems. The small run time improvement shown in Fig. \ref{fig:benchmarksGCE} for the P100 machine with 16 cores comes with more than a factor of four increase in financial cost per hour. 
Thus, a K80 GPU or optimized MKL FFTs running on a multi-core desktop machine (such as the iMac Pro in Fig. \ref{fig:MacPro}) is likely to provide more simulations per dollar.

These approaches have preserved existing POPPY numerical accuracy; however, even in extreme contrast regimes, numerical approximations or single precision calculations may provide further efficiency gains without significant loss of model accuracy. 
For example, the exponential function, the primary driver of run time of the model system when using a high-end-GPU, can be accelerated by interpolation or bit manipulation\cite{yamamoto_computational_2004}.
For certain cases, sparse-\gls{FFT}s\cite{hassanieh_simple_2012} may provide additional gains, but these algorithms typically discard weak signals and they would require great care to ensure accuracy, particularly for coronagraph modeling where signals regularly span ten or more orders orders of magnitude.
For highly specialized calculations, entirely GPU based calculations of diffraction\cite{shimobaba_numerical_2008} provide the most promising avenue for improved performance. However, the results herein show significant run time gains can be made by leveraging optimized computational hardware and libraries within a user-friendly python scientific computing framework. 
\section{Summary}
For a realistic 10 plane optical system:
\begin{itemize}
\item An approximately $3\times$ speed-up was realized using optimized math libraries and a 5$\times$ speed up was found using research-grade GPUs  for some calculations.
\item NumExpr library’s chunked, multicore array math provides significant speed gains.
\item FFTW and Intel MKL-FFT provide large advantages over built in NumPy FFT functions.
\item Pyculib GPU FFT and Numba-compiled FFTSHIFT functions provide large speedup on research grade GPUs for large arrays.
\item A consumer grade GPU provided negligible speed-up for FFT only calculations using OpenCL.
\end{itemize}

\acknowledgments 
The authors would  like to thank all of the contributors to the POPPY library for their inputs and assistance in its development, in particular Joseph Long, Neil Zimmerman, Maciek Grochowicz, and Phillip Springer for their contributions to the Fresnel propagation implementation and WFIRST SPC example model. 
Additional thanks to Ludwig Schmidt and Haitham Hassanieh  for their advice on optimizing \gls{FFT} run times. 

Portions of this work were supported by the James Webb Space Telescope program.
JWST is a joint project of the U.S. National Aeronautics and Space Administration (NASA), the European Space Agency, and the Canadian Space Agency. The Space Telescope Science Institute is operated by the Association of Universities for Research in Astronomy, Inc., under NASA contracts NAS5-26555 and  NAS5-03127.
Portions of this work were supported by the WFIRST Science Investigation Team, NASA award NNG16PJ24C. 
This research made use of community-developed core Python packages, including: Astropy\cite{the_astropy_collaboration_astropy_2013}, Matplotlib\cite{hunter_matplotlib_2007}, SciPy\cite{jones_scipy_2001}, 
the IPython Interactive Computing architecture \cite{perez_ipython_2007}, and Jupyter\cite{kluyver_jupyter_2016}.


\newpage
\appendix

\lstset{basicstyle = \verbatim@font,
	language=Python,
	breaklines=True,
 	 keepspaces=true,                 
 	 columns=flexible
}

\section{Talbot Effect Example Code}
The following POPPY prescription performs the optical propagation shown in Fig. \ref{fig:talbot}.

\begin{lstlisting}[frame=single]  % Start your code-block with one blank line

import poppy
import astropy.units as u

wf_f = poppy.FresnelWavefront(beam_radius=2*u.cm,wavelength=0.5*u.um,
                              npix=256,oversample=8)

sineWFE = poppy.wfe.SineWaveWFE(spatialfreq=500,amplitude=5e-9)
wf_f* = sineWFE
wf_f *= poppy.CircularAperture(radius=wf_f.diam/2)

Z_t = 2*((1/sineWFE.sine_spatial_freq))**2/wf_f.wavelength
wf_f.propagate_fresnel(Z_t/4.)
\end{lstlisting}\label{code:talbot}

\section{Example System}\label{code:fresnel}
The benchmark plane-to-plane, system is a simplified model of the \gls{WFIRST} Shaped Pupil Bow-Tie Coronagraph\cite{zimmerman_shaped_2016}, adapted from a model of CGI included in WebbPSF.
Note, this is not intended as a precise model of WFIRST CGI but rather illustrates the building blocks of a a coronagraph system and Fresnel propagation. It makes use of data files on optical elements available in the WebbPSF reference data (\url{http://www.stsci.edu/~mperrin/software/webbpsf/webbpsf-data-0.7.0.tar.gz}).
\begin{small}
\begin{lstlisting}[frame=single]  % Start your code-block with one blank line

import os
#export environment variable:
os.environ['WEBBPSF_PATH'] = os.path.expanduser('~/STScI/WFIRST/webbpsf-data')

import poppy
import astropy.units as u

def WFIRSTSPC(npix=256,ratio=0.25):
    Tel_fname = os.path.join(os.environ['WEBBPSF_PATH'], "AFTA_CGI_C5_Pupil_onax_256px_flip.fits")
    SP_fname = os.path.join(os.environ['WEBBPSF_PATH'], "CGI/optics/CHARSPC_SP_256pix.fits.gz")
    FPM_fname = os.path.join(os.environ['WEBBPSF_PATH'], "CGI/optics/CHARSPC_FPM_25WA90_2x65deg_-_FP1res4_evensamp_D072_F770.fits.gz")
    LS_fname = os.path.join(os.environ['WEBBPSF_PATH'], "CGI/optics/SPC_LS_30D88_256pix.fits.gz")


    D_prim = 2.37 * u.m
    D_relay = 20 * u.mm
    fr_pri = 7.8
    fl_pri = D_prim * fr_pri
    fl_m2 = fl_pri * D_relay / D_prim
    fr_m3 = 20.
    fl_m3 = fr_m3 * D_relay



    oversamp=4
    wfirst_optsys = poppy.FresnelOpticalSystem(pupil_diameter=D_prim, 
                                               beam_ratio=ratio,
                                               npix=npix)

    telap = poppy.FITSOpticalElement(transmission=Tel_fname)
    SP = poppy.FITSOpticalElement(transmission=SP_fname)

    #default FPM pixelscale is in arcsecs
    FPM = poppy.FITSOpticalElement(transmission=FPM_fname,
                                   planetype=poppy.poppy_core.PlaneType.intermediate,
                                   pixelscale=0.005)
    SP.pixelscale=0.5*u.cm/SP.shape[0]/u.pix
    FPM.pixelscale=0.5*u.cm/SP.shape[0]/u.pix
    
    m1 = poppy.QuadraticLens(fl_pri, name='Primary')
    m2 = poppy.QuadraticLens(fl_m2, name='M2')
    m3 = poppy.QuadraticLens(fl_m3, name='M3')
    m4 = poppy.QuadraticLens(fl_m3, name='M4')
    m5 = poppy.QuadraticLens(fl_m3, name='M5')
    m6 = poppy.QuadraticLens(fl_m3, name='M6')

    wfirst_optsys.add_optic(telap)
    wfirst_optsys.add_optic(m1)
    wfirst_optsys.add_optic(m2, distance = fl_pri + fl_m2)
    wfirst_optsys.add_optic(m3, distance = fl_m2 + fl_m3)
    wfirst_optsys.add_optic(m4, distance = 2*fl_m3)
    wfirst_optsys.add_optic(SP, distance = fl_m3)
    wfirst_optsys.add_optic(m5, distance = fl_m3)
    wfirst_optsys.add_optic(FPM, distance = fl_m3)
    wfirst_optsys.add_optic(m5, distance = 2*fl_m3)
    wfirst_optsys.add_optic(poppy.ScalarTransmission(
                               planetype=poppy.poppy_core.PlaneType.intermediate,
                               name='focus',),
                            distance=fl_m3+0.39999923*u.m)
    return wfirst_optsys
    
 psf = WFIRSTSPC(npix=1025).calc_psf()


\end{lstlisting}\label{code:examplesystem}
\end{small}

\newpage
\bibliography{poppy2018spieDontEdit,references} 

\begin{thebibliography}{10}

\bibitem{perrin_simulating_2012}
Perrin, M.~D., Soummer, R., Elliott, E.~M., Lallo, M.~D., and Sivaramakrishnan,
  A., ``Simulating point spread functions for the {{James Webb Space
  Telescope}} with {{WebbPSF}},'' in [{\em Proc.
  {{SPIE}}}{\nolinebreak\hspace{0.1em}]},   {\bf 8442},  84423D--84423D--11
  (2012).
\newblock \url{http://dx.doi.org/10.1117/12.925230}.

\bibitem{perrin_poppy_2016}
Perrin, M., Long, J., Douglas, E., Sivaramakrishnan, A., Slocum, C., and
  {others}, ``{{POPPY}}: {{Physical Optics Propagation}} in {{PYthon}},'' {\em
  Astrophysics Source Code Library} ,  ascl:1602.018 (Feb. 2016).
\newblock \url{http://adsabs.harvard.edu/abs/2016ascl.soft02018P}.

\bibitem{the_astropy_collaboration_astropy_2013}
{The Astropy Collaboration}, Robitaille, T.~P., Tollerud, E.~J., Greenfield,
  P., Droettboom, M., Bray, E., Aldcroft, T., Davis, M., Ginsburg, A.,
  {Price-Whelan}, A.~M., Kerzendorf, W.~E., Conley, A., Crighton, N., Barbary,
  K., Muna, D., Ferguson, H., Grollier, F., Parikh, M.~M., Nair, P.~H.,
  G{\"u}nther, H.~M., Deil, C., Woillez, J., Conseil, S., Kramer, R., Turner,
  J. E.~H., Singer, L., Fox, R., Weaver, B.~A., Zabalza, V., Edwards, Z.~I.,
  Azalee~Bostroem, K., Burke, D.~J., Casey, A.~R., Crawford, S.~M., Dencheva,
  N., Ely, J., Jenness, T., Labrie, K., Lim, P.~L., Pierfederici, F., Pontzen,
  A., Ptak, A., Refsdal, B., Servillat, M., and Streicher, O., ``Astropy: {{A}}
  community {{Python}} package for astronomy,'' {\em Astronomy \&
  Astrophysics}~{\bf 558},  A33 (Oct. 2013).
\newblock \url{http://www.aanda.org/10.1051/0004-6361/201322068}.

\bibitem{goodman_introduction_2005}
Goodman, J.~W.,  [{\em Introduction to {{Fourier
  Optics}}}{\nolinebreak\hspace{0.1em}]}, {Roberts and Company Publishers}
  (2005).

\bibitem{soummer_fast_2007}
Soummer, R., Pueyo, L., Sivaramakrishnan, A., and Vanderbei, R.~J., ``Fast
  computation of {{Lyot}}-style coronagraph propagation,'' {\em Opt.
  Express}~{\bf 15},  15935--15951 (Nov. 2007).
\newblock \url{http://www.opticsexpress.org/abstract.cfm?URI=oe-15-24-15935}.

\bibitem{lawrence_optical_1992}
Lawrence, G.~N., ``Optical {{Modeling}},'' in [{\em Applied {{Optics}} and
  {{Optical Engineering}}.}{\nolinebreak\hspace{0.1em}]},  Shannon, R.~R. and
  Wyant., J.~C., eds.,  {\bf XI}, {Academic Press}, New York (1992).

\bibitem{krist_proper_2007}
Krist, J.~E., ``{{PROPER}}: An optical propagation library for {{IDL}},'' in
  [{\em Proc. {{SPIE}}}{\nolinebreak\hspace{0.1em}]},   {\bf 6675},
  66750P--66750P--9 (2007).
\newblock \url{http://dx.doi.org/10.1117/12.731179}.

\bibitem{marois_end--end_2008}
Marois, C., Macintosh, B., Soummer, R., Poyneer, L., and Bauman, B., ``An
  end-to-end polychromatic {{Fresnel}} propagation model of {{GPI}},''  {\bf
  7015},  70151T (July 2008).
\newblock \url{http://adsabs.harvard.edu/abs/2008SPIE.7015E..1TM}.

\bibitem{yaitskova_foros_2010}
Yaitskova, N., Dohlen, K., Rabou, P., Boccaletti, A., Carbillet, M., Beuzit,
  J.-L., Kasper, M., and Hubin, N., ``{{FOROS}}: {{Fresnel}} optical
  propagation code for {{SPHERE}},''  {\bf 7735},  77352T--77352T--13 (2010).
\newblock \url{http://dx.doi.org/10.1117/12.856253}.

\bibitem{krist_assessing_2011}
Krist, J.~E., Belikov, R., Pueyo, L., Mawet, D.~P., Moody, D., Trauger, J.~T.,
  and Shaklan, S.~B., ``Assessing the performance limits of internal
  coronagraphs through end-to-end modeling: A {{NASA TDEM}} study,'' in [{\em
  Proc. {{SPIE}}}{\nolinebreak\hspace{0.1em}]},   {\bf 8151},
  81510E--81510E--16 (2011).
\newblock \url{http://dx.doi.org/10.1117/12.892772}.

\bibitem{douglas_end--end_2015}
Douglas, E.~S., Hewasawam, K., Mendillo, C.~B., Cahoy, K.~L., Cook, T.~A.,
  Finn, S.~C., Howe, G.~A., Kuchner, M.~J., Lewis, N.~K., Marinan, A.~D.,
  Mawet, D., and Chakrabarti, S., ``End-to-end simulation of high-contrast
  imaging systems: Methods and results for the {{PICTURE}} mission family,'' in
  [{\em Proc. {{SPIE}}}{\nolinebreak\hspace{0.1em}]},   {\bf 9605},
  96051A--96051A--13 (2015).
\newblock \url{http://dx.doi.org/10.1117/12.2187262}.

\bibitem{mendillo_optical_2017}
Mendillo, C.~B., Howe, G.~A., Hewawasam, K., Martel, J., Finn, S.~C., Cook,
  T.~A., and Chakrabarti, S., ``Optical tolerances for the {{PICTURE}}-{{C}}
  mission: Error budget for electric field conjugation, beam walk, surface
  scatter, and polarization aberration,'' in [{\em Proc
  {{SPIE}}}{\nolinebreak\hspace{0.1em}]},   {\bf 10400},  1040010,
  {International Society for Optics and Photonics} (Sept. 2017).
\newblock
  \url{https://www.spiedigitallibrary.org/conference-proceedings-of-spie/10400/1040010/Optical-tolerances-for-the-PICTURE-C-mission--error-budget/10.1117/12.2274105.short}.

\bibitem{lumbres_modeling_2018}
Lumbres, J. and {et al.}, ``Modeling coronagraphic extreme wavefront control
  systems for high contrast imaging in ground and space telescope missions,''
  in [{\em Proc {{SPIE}}}{\nolinebreak\hspace{0.1em}]},  {SPIE}, Austin Texas
  (June 2018).

\bibitem{ndiaye_high-contrast_2013}
N'Diaye, M., Choquet, E., Pueyo, L., Elliot, E., Perrin, M.~D., Wallace, J.~K.,
  Groff, T., Carlotti, A., Mawet, D., Sheckells, M., Shaklan, S., Macintosh,
  B., Kasdin, N.~J., and Soummer, R., ``High-contrast imager for complex
  aperture telescopes ({{HiCAT}}): 1. testbed design,'' in [{\em Proc
  {{SPIE}}}{\nolinebreak\hspace{0.1em}]},   {\bf 8864},  88641K--88641K--10
  (2013).
\newblock \url{http://dx.doi.org/10.1117/12.2023718}.

\bibitem{noecker_coronagraph_2016}
Noecker, M.~C., Zhao, F., Demers, R., Trauger, J., Guyon, O., and Kasdin,
  N.~J., ``Coronagraph instrument for {{WFIRST}}-{{AFTA}},'' {\em JATIS,
  JATIAG}~{\bf 2},  011001 (Mar. 2016).
\newblock
  \url{https://www.spiedigitallibrary.org/journals/Journal-of-Astronomical-Telescopes-Instruments-and-Systems/volume-2/issue-1/011001/Coronagraph-instrument-for-WFIRST-AFTA/10.1117/1.JATIS.2.1.011001.short}.

\bibitem{douglas_wavefront_2018}
Douglas, E.~S., Mendillo, C.~B., Cook, T.~A., Cahoy, K.~L., and Chakrabarti,
  S., ``Wavefront sensing in space: Flight demonstration {{II}} of the
  {{PICTURE}} sounding rocket payload,'' {\em JATIS, JATIAG}~{\bf 4},  019003
  (Mar. 2018).
\newblock
  \url{https://www.spiedigitallibrary.org/journals/Journal-of-Astronomical-Telescopes-Instruments-and-Systems/volume-4/issue-1/019003/Wavefront-sensing-in-space--flight-demonstration-II-of-the/10.1117/1.JATIS.4.1.019003.short}.

\bibitem{krist_technology_2013}
Krist, J.~E., Belikov, R., Mawet, D.~P., Moody, D., Pueyo, L., Shaklan, S.~B.,
  and Trauger, J.~T., ``Technology {{Milestone}} 1 {{Results Report}}:
  {{Assessing}} the performance limits of internal coronagraphs through
  end-to-end modeling,'' Tech. Rep. JPL Document D-74425, {JPL} (2013).
\newblock \url{available\%0020from\%0020exep.jpl.nasa.gov/technology}.

\bibitem{macintosh_gemini_2008}
Macintosh, B.~A., Graham, J.~R., Palmer, D.~W., Doyon, R., Dunn, J., Gavel,
  D.~T., Larkin, J., Oppenheimer, B., Saddlemyer, L., and Sivaramakrishnan, A.,
  ``The {{Gemini Planet Imager}}: From science to design to construction,'' in
  [{\em Proc. {{SPIE}}}{\nolinebreak\hspace{0.1em}]},   {\bf 7015},  701518
  (2008).
\newblock \url{http://144.206.159.178/ft/CONF/16417791/16417811.pdf}.

\bibitem{morgan_initial_2015}
Morgan, R. and Siegler, N., ``Initial look at the coronagraph technology gaps
  for direct imaging of exo-earths,'' in [{\em Proc.
  {{SPIE}}}{\nolinebreak\hspace{0.1em}]},   {\bf 9605},  96052I (Sept. 2015).
\newblock \url{http://adsabs.harvard.edu/abs/2015SPIE.9605E..2IM}.

\bibitem{bolcar_large_2017}
Bolcar, M.~R., ``The {{Large UV}}/{{Optical}}/{{Infrared}} ({{LUVOIR}})
  surveyor: {{Decadal Mission}} concept technology development overview,'' in
  [{\em Proc {{SPIE}}}{\nolinebreak\hspace{0.1em}]},   {\bf 10398},  103980A,
  {International Society for Optics and Photonics} (Sept. 2017).
\newblock
  \url{https://www.spiedigitallibrary.org/conference-proceedings-of-spie/10398/103980A/The-Large-UV-Optical-Infrared-LUVOIR-surveyor--Decadal-Mission/10.1117/12.2273853.short}.

\bibitem{cooley_1965}
Cooley, J.~W. and Tukey, J.~W., ``{An algorithm for the machine calculation of
  complex Fourier series},'' {\em Mathematics of Computation}~{\bf 19},
  297--301 (1965).
\newblock URL: {\tt http://cr.yp.to/\allowbreak bib/\allowbreak
  entries.html\#\allowbreak 1965/\allowbreak cooley}.

\bibitem{jones_scipy_2001}
Jones, E., Oliphant, T., and Peterson, P., ``{{SciPy}}: {{Open}} source
  scientific tools for {{Python}},'' {\em http://www. scipy. org/}  (2001).
\newblock \url{http://www.citeulike.org/group/2018/article/2644428}.

\bibitem{fangohr_comparison_2004}
Fangohr, H., ``A {{Comparison}} of {{C}}, {{MATLAB}}, and {{Python}} as
  {{Teaching Languages}} in {{Engineering}},'' in [{\em Computational
  {{Science}} - {{ICCS}} 2004}{\nolinebreak\hspace{0.1em}]},  {\em Lecture
  Notes in Computer Science},  1210--1217, {Springer, Berlin, Heidelberg} (June
  2004).
\newblock
  \url{https://link.springer.com/chapter/10.1007/978-3-540-25944-2_157}.

\bibitem{greenfield_reaching_2007}
Greenfield, P., ``Reaching for the {{Stars}} with {{Python}},'' {\em Computing
  in Science \& Engineering}~{\bf 9}(3),  38--40 (2007).
\newblock \url{http://ieeexplore.ieee.org/document/4160254/}.

\bibitem{hirst_ureka_2014-1}
Hirst, P., Slocum, C., Turner, J., Sienkiewicz, M., Greenfield, P., Hogan, E.,
  Simpson, M., and Labrie, K., ``Ureka: {{A Distribution}} of {{Python}} and
  {{IRAF Software}} for {{Astronomy}},''  {\bf 485},  335 (May 2014).
\newblock \url{http://adsabs.harvard.edu/abs/2014ASPC..485..335H}.

\bibitem{Greenfield_2003}
Greenfield, P., Miller, T., Hsu, J.-C., and L~White, R., ``numarray: A new
  scientific array package for python,'' in [{\em Medical Imaging:\ Image
  Processing}{\nolinebreak\hspace{0.1em}]},  {\em Proceedings of the O'Reilly
  Open Source Convention} (5 2003).

\bibitem{_globalinterpreterlock_2018}
``{{GlobalInterpreterLock}} - {{Python Wiki}}.''
  \url{https://wiki.python.org/moin/GlobalInterpreterLock} (2018).

\bibitem{cooke_numexpr_2018}
Cooke, D.~M. and Alted, F., ``Numexpr: {{Fast}} numerical array expression
  evaluator for {{Python}}, {{NumPy}}, {{PyTables}}, pandas, bcolz and more.''
  \url{https://github.com/pydata/numexpr} (May 2018).

\bibitem{lam_numba_2015}
Lam, S.~K., Pitrou, A., and Seibert, S., ``Numba: {{A LLVM}}-based {{Python JIT
  Compiler}},'' in [{\em Proceedings of the {{Second Workshop}} on the {{LLVM
  Compiler Infrastructure}} in {{HPC}}}{\nolinebreak\hspace{0.1em}]},  {\em
  LLVM '15},  7:1--7:6, {ACM}, New York, NY, USA (2015).
\newblock \url{http://doi.acm.org/10.1145/2833157.2833162}.

\bibitem{nickolls_scalable_2008}
Nickolls, J., Buck, I., Garland, M., and Skadron, K., ``Scalable {{Parallel
  Programming}} with {{CUDA}},'' in [{\em {{ACM SIGGRAPH}} 2008
  {{Classes}}}{\nolinebreak\hspace{0.1em}]},  {\em SIGGRAPH '08},  16:1--16:14,
  {ACM}, New York, NY, USA (2008).
\newblock \url{http://doi.acm.org/10.1145/1401132.1401152}.

\bibitem{stone_opencl_2010}
Stone, J.~E., Gohara, D., and Shi, G., ``{{OpenCL}}: {{A Parallel Programming
  Standard}} for {{Heterogeneous Computing Systems}},'' {\em Comput Sci
  Eng}~{\bf 12},  66--72 (May 2010).
\newblock \url{https://www.ncbi.nlm.nih.gov/pmc/articles/PMC2964860/}.

\bibitem{shimobaba_numerical_2008}
Shimobaba, T., Ito, T., Masuda, N., Abe, Y., Ichihashi, Y., {Hirotaka
  Nakayama}, Takada, N., Shiraki, A., and Sugie, T., ``Numerical calculation
  library for diffraction integrals using the graphic processing unit: The
  {{GPU}}-based wave optics library,'' {\em J. Opt. A: Pure Appl. Opt.}~{\bf
  10}(7),  075308 (2008).
\newblock \url{http://stacks.iop.org/1464-4258/10/i=7/a=075308}.

\bibitem{pavlyk_accelerating_2017}
Pavlyk, O., Nagorny, D., {Guzman-Ballen}, A., Malakhov, A., Liu, H., Totoni,
  E., Anderson, A., and Maidanov, S., ``Accelerating {{Scientific Python}} with
  {{Intel Optimizations}},'' {\em PROC. OF THE 16th PYTHON IN SCIENCE CONF} ,
  7 (2017).
\newblock
  \url{https://conference.scipy.org/proceedings/scipy2017/pdfs/oleksandr_pavlyk.pdf}.

\bibitem{frigo_design_2005}
Frigo, M. and Johnson, S.~G., ``The {{Design}} and {{Implementation}} of
  {{FFTW3}},'' {\em Proceedings of the IEEE}~{\bf 93},  216--231 (Feb. 2005).

\bibitem{steinbach_gearshifft_2017}
Steinbach, P. and Werner, M., ``Gearshifft - {{The FFT Benchmark Suite}} for
  {{Heterogeneous Platforms}},'' {\em arXiv:1702.00629 [cs]}~{\bf 10266}
  (2017).
\newblock \url{http://arxiv.org/abs/1702.00629}.

\bibitem{_pyculib_2018}
``Pyculib: {{Pyculib}} - {{Python}} bindings for {{CUDA}} libraries.''
  \url{https://github.com/numba/pyculib} (May 2018).

\bibitem{_anaconda_}
``Anaconda {{Accelerate}} | {{Anaconda}}: {{Documentation}}.''
  \url{https://docs.anaconda.com/accelerate/}.

\bibitem{_cufft_2018}
``{{cuFFT}} | {{NVIDIA}}.'' \url{https://developer.nvidia.com/cufft} (May
  2018).

\bibitem{_clfft_2018}
``{{clFFT}}: A software library containing {{FFT}} functions written in
  {{OpenCL}}.'' \url{https://github.com/clMathLibraries/clFFT} (June 2018).

\bibitem{_intel_-1}
``{{Intel}}\textregistered{} {{Distribution}} for {{Python}}* |
  {{Intel}}\textregistered{} {{Software}}.''
  \url{https://software.intel.com/en-us/distribution-for-python}.

\bibitem{akeret_hope_2015}
Akeret, J., Gamper, L., Amara, A., and Refregier, A., ``{{HOPE}}: {{A Python}}
  just-in-time compiler for astrophysical computations,'' {\em Astronomy and
  Computing}~{\bf 10},  1--8 (Apr. 2015).
\newblock
  \url{http://www.sciencedirect.com/science/article/pii/S2213133714000687}.

\bibitem{abdellah_cufftshift_2014}
Abdellah, M., ``{{CufftShift}}: {{High}} performance {{CUDA}}-accelerated
  {{FFT}}-shift library,'' in [{\em Simulation
  {{Series}}}{\nolinebreak\hspace{0.1em}]},   {\bf 46} (Apr. 2014).

\bibitem{douglas_douglase/poppy_benchmarking_2018}
Douglas, E.~S., ``Douglase/poppy\_benchmarking: {{Initial}} release of notebook
  for {{SPIE AST}} 2018.'' \url{https://zenodo.org/record/1286650} (June 2018).

\bibitem{yamamoto_computational_2004}
Yamamoto, A., Kitamura, Y., and Yamane, Y., ``Computational efficiencies of
  approximated exponential functions for transport calculations of the
  characteristics method,'' {\em Annals of Nuclear Energy}~{\bf 31},
  1027--1037 (June 2004).
\newblock
  \url{http://www.sciencedirect.com/science/article/pii/S0306454904000234}.

\bibitem{hassanieh_simple_2012}
Hassanieh, H., Indyk, P., Katabi, D., and Price, E., ``Simple and {{Practical
  Algorithm}} for {{Sparse Fourier Transform}},'' in [{\em Proceedings of the
  {{Twenty}}-Third {{Annual ACM}}-{{SIAM Symposium}} on {{Discrete
  Algorithms}}}{\nolinebreak\hspace{0.1em}]},  {\em SODA '12},  1183--1194,
  {Society for Industrial and Applied Mathematics}, Philadelphia, PA, USA
  (2012).
\newblock \url{http://dl.acm.org/citation.cfm?id=2095116.2095209}.

\bibitem{hunter_matplotlib_2007}
Hunter, J.~D., ``Matplotlib: {{A 2D}} graphics environment,'' {\em Computing In
  Science \& Engineering}~{\bf 9}(3),  90--95 (2007).

\bibitem{perez_ipython_2007}
P{\'e}rez, F. and Granger, B., ``{{IPython}}: {{A System}} for {{Interactive
  Scientific Computing}},'' {\em Computing in Science Engineering}~{\bf 9},
  21--29 (May 2007).

\bibitem{kluyver_jupyter_2016}
Kluyver, T., {Ragan-Kelley}, B., P{\'e}rez, F., Granger, B.~E., Bussonnier, M.,
  Frederic, J., Kelley, K., Hamrick, J.~B., Grout, J., and Corlay, S.,
  ``Jupyter {{Notebooks}}-a publishing format for reproducible computational
  workflows.,'' in [{\em {{ELPUB}}}{\nolinebreak\hspace{0.1em}]},   87--90
  (2016).

\bibitem{zimmerman_shaped_2016}
Zimmerman, N.~T., Eldorado~Riggs, A.~J., Jeremy~Kasdin, N., Carlotti, A., and
  Vanderbei, R.~J., ``Shaped pupil {{Lyot}} coronagraphs: High-contrast
  solutions for restricted focal planes,'' {\em J. Astron. Telesc. Instrum.
  Syst}~{\bf 2}(1),  011012--011012 (2016).
\newblock \url{http://dx.doi.org/10.1117/1.JATIS.2.1.011012}.

\end{thebibliography}
\bibliographystyle{spiebib} 

\end{document}